\documentclass[sigconf]{acmart}

\AtBeginDocument{%
  \providecommand\BibTeX{{%
    \normalfont B\kern-0.5em{\scshape i\kern-0.25em b}\kern-0.8em\TeX}}}

\usepackage[ruled, linesnumbered]{algorithm2e}
\usepackage{xspace}
\usepackage{subcaption}
\newcommand{\ms}{\ensuremath{\mathsf{MetaSelector}}\xspace}
\newcommand{\calM}{\ensuremath{\mathcal M}}
\newcommand{\calS}{\ensuremath{\mathcal S}}

\setcopyright{acmcopyright}

\copyrightyear{2020}
\acmYear{2020}
\setcopyright{iw3c2w3}
\acmConference[WWW '20]{Proceedings of The Web Conference 2020}{April 20--24, 2020}{Taipei, Taiwan}
\acmBooktitle{Proceedings of The Web Conference 2020 (WWW '20), April 20--24, 2020, Taipei, Taiwan}
\acmPrice{}
\acmDOI{10.1145/3366423.3379999}
\acmISBN{978-1-4503-7023-3/20/04}

\begin{document}

\title{MetaSelector: Meta-Learning for Recommendation with User-Level Adaptive Model Selection}


\author{Mi Luo}
\affiliation{\institution{Huawei Noah's Ark Lab}}
\email{rosemaryluo@outlook.com}
\authornote{Equal contribution. This work was done when Mi Luo was intern at Huawei Noah's Ark Lab.}

\author{Fei Chen}
\authornotemark[1]
\affiliation{\institution{Huawei Noah's Ark Lab}}
\email{chen.f@huawei.com}

\author{Pengxiang Cheng}
\affiliation{\institution{Huawei Noah's Ark Lab}}
\email{chengpengxiang1@huawei.com}

\author{Zhenhua Dong}
\affiliation{\institution{Huawei Noah's Ark Lab}}
\email{dongzhenhua@huawei.com}

\author{Xiuqiang He}
\affiliation{\institution{Huawei Noah's Ark Lab}}
\email{hexiuqiang1@huawei.com}

\author{Jiashi Feng}
\affiliation{\institution{National University of Singapore}}
\email{elefjia@nus.edu.sg}

\author{Zhenguo Li}
\affiliation{\institution{Huawei Noah's Ark Lab}}
\email{li.zhenguo@huawei.com}

\renewcommand{\shortauthors}{Luo and Chen, et al.}

\begin{abstract}
	Recommender systems often face heterogeneous datasets containing highly personalized historical data of users, where no single model could give the best recommendation for every user. We observe this ubiquitous phenomenon on both public and private datasets and address the model selection problem in pursuit of optimizing the quality of recommendation for each user. We propose a meta-learning framework to facilitate user-level adaptive model selection in recommender systems. In this framework, a collection of recommenders is trained with data from all users, on top of which a model selector is trained via meta-learning to select the best single model for each user with the user-specific historical data. We conduct extensive experiments on two public datasets and a real-world production dataset, demonstrating that our proposed framework achieves improvements over single model baselines and sample-level model selector in terms of AUC and LogLoss. In particular, the improvements may lead to huge profit gain when deployed in online recommender systems.
\end{abstract}



\maketitle

\section{Introduction}
\label{sec:intro}

In recommender systems, deep learning has played an increasingly important role in discovering useful behavior patterns from huge amount of user data and providing precise and personalized recommendation in various scenarios~\cite{wang2015collaborative, cheng2016wide,he2017neural, wang2019neural}. Data from one user may be sparse and insufficient to support effective model training.
In practice, deep neural networks are trained collaboratively on a large number of users, and it is important to distinguish the specific users to make personalized recommendation.
Certain user identification processes are therefore often performed in alignment with the model training procedure, such as encoding a unique ID or user history information for each user\cite{zhou2018deep}, or fine-tuning the recommender on user local data before making recommendations~\cite{chen2018federated}.

Although certain recommendation models could achieve better overall performance than other models, it is unlikely that there is a single model that performs better than other models for every user~\cite{ekstrand2015letting, ekstrand2012when}. In other words, the best performance on different users may be achieved by different recommendation models. We observed this phenomenon on both private production and public datasets. For instance, in an online advertising system, multiple CTR prediction models are deployed simultaneously~\cite{zhou2018deep}. We found that no single model performs best on all users. Moreover, in terms of averaged evaluation, no single model achieves the all-time best performance.
This implies that the performance of recommendation models is sensitive to user-specific data. Consequently, user-level model design in deep recommender systems is of both research interests and practical values.

In this work, we address the problem of user-level model selection to improve personalized recommendation quality. Given a collection of deep models, the goal is to select the best model from them for each individual user or to combine these models to maximize their strengths. We introduce a model selector on top of specific recommendation models to decide which model to use for an user. Considering the fast adaptation ability of the recently revived meta-learning, we formulate the model selection problem under the meta-learning setting and propose \ms which trains the model selector and the recommendation models via the meta-learning methodology~\cite{schmidhuber1987evolutionary,thrun2012learning,andrychowicz2016learning,vinyals2016matching,ravi2017optimization,finn2017model,huang2019meta}.

Meta-learning algorithms learn to efficiently solve new tasks by extracting prior information from a number of related tasks. Of particular interest are optimization-based approaches, such as the popular Model-Agnostic Meta-Learning (MAML) algorithm~\cite{finn2017model}, that apply to a wide range of models whose parameters are updated by stochastic gradient descent (SGD), with little requirement on the model structure. MAML involves a bi-level meta-learning process. The outer loop is on task level, where the algorithm maintains an \emph{initialization} for the parameters. The objective is to optimize the initialization such that when applied to a new task, the initialization leads to optimal performance on the test set after one or a few gradient updates on the training set. The inner loop is on sample level and executed within tasks. Receiving the initialization maintained in the outer loop, the algorithm adapts parameters on the support (training) set and evaluates the model on the query (test) set. The evaluation result on test set returns a loss signal to the outer loop. After meta-training, in the meta-testing or deployment phase the learned initialization enables fast adaptation on new tasks.

\begin{figure}[t]
    \centering
    \includegraphics[width=0.4\textwidth]{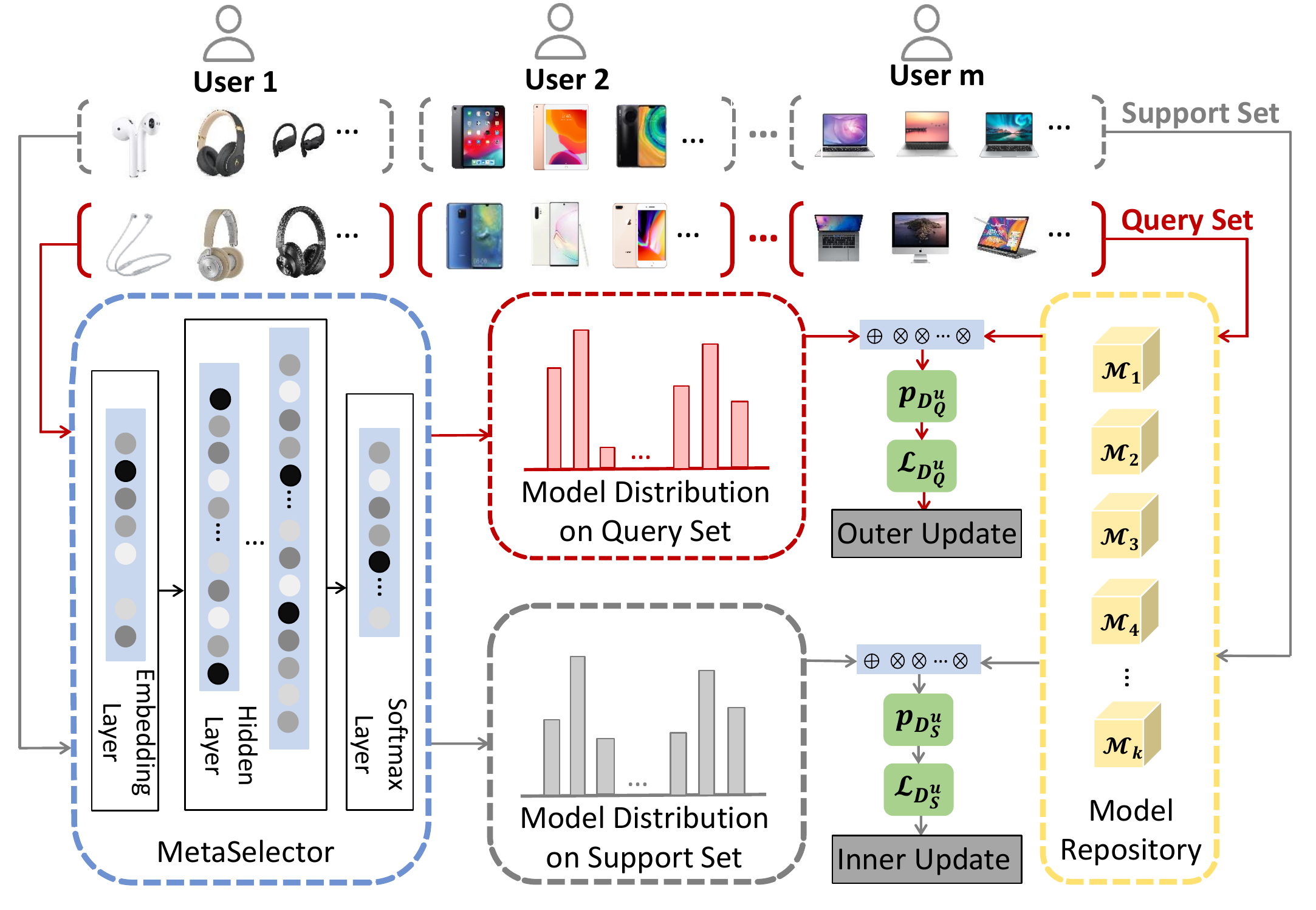}
    \vspace{-10pt}
    \caption{The \ms framework.}
    \label{fig:workflow}
    \vspace{-20pt}
\end{figure}

Mete-Learning is well-suited for model selection if we regard each task as learning to predict user preference for
selecting models. As shown in Figure~\ref{fig:workflow}, in our method, we use optimization-based meta-learning methods to construct \ms that learns to make model selection from a number of tasks, where a task consists of data from one user. Given a recommendation request as input, \ms outputs a probability distribution over the recommendation models. In the meta-training phase, an initialization for \ms is optimized through episodic learning~\cite{finn2017model}. In each episode, a batch of tasks are sampled, each with a support set and a query set. On the support set of each task, a soft model selection is made based on the output of \ms. 
The parameters of \ms are updated using the training loss obtained by comparing the final prediction with ground truth. Then the adapted \ms is evaluated on the query set, and test loss is similarly computed to update the initialization in the outer loop. The recommendation models are updated together in the outer loop, which can be optionally pre-trained before the meta-training process. In the deployment phase, with the learned initialization, \ms adapts to individual users using personalized historical data (support sets), and aggregates results of recommendation models for new queries.

We experimentally demonstrate effectiveness of our proposed method on two public datasets and a production dataset. In all experiments, \ms significantly improves over baseline models in terms of AUC and LogLoss, indicating that \ms can effectively weigh towards better models at the user level. 
We also observe that pre-training the recommendation models is crucial to express the power of \ms.

\textbf{Contributions.} To summarize, our contributions are three-fold. Firstly, we address the problem of model selection for recommender systems, motivated by the observation of varying performance of different models among users on public and production datasets. Secondly, we propose a novel framework \ms which introduces meta-learning to formulate a user-level model selection module in a hybrid recommender system that involves the combination of two or more recommendation models. This framework can be trained end-to-end and requires no manual definition of meta-features. To the best of our knowledge, this is the first work to study recommendation model selection problem from the optimization-based meta-learning perspective. Thirdly, we run extensive experiments on both public and private production datasets to provide the insight into which level to optimize in model selection. The results indicate that \ms can improve the performance over single model baseline and sample-level selector,  showing the potential of \ms in real-world recommender systems.

\section{RELATED WORK}
\label{sec:releatedwork}
Since we study how to apply meta-learning for model selection in a hybrid recommender system, we first survey relevant work on meta-learning and model selection.
Besides, we initially observed the varying performances of recommendation models in a real-world industrial CTR prediction problem. Hence we also review some classic CTR prediction models. 

\subsection{Optimization-Based Meta-Learning}
In meta-learning, or ``learning to learn'', the goal is to learn a model on a collection of tasks, such that it can achieve fast adaptation to new tasks \cite{chen2019closer}. One research direction is metric-based meta-learning, aiming to learn the similarity between samples within tasks. Representative works include Matching Network~\cite{vinyals2016matching} and Prototypical Networks~\cite{snell2017prototypical}. Another promising direction is optimization-based meta-learning which has recently demonstrated effectiveness on few-shot classification problems by ``learning to fine-tune''. Among the various methods, some focus on learning an optimizer such as the LSTM-based meta-learner \cite{ravi2017optimization} and the Meta Networks with an external memory \cite{munkhdalai2017meta}. Another research branch aims to learn a good model initialization~\cite{finn2017model,li2017meta,nichol2018reptile}, such that the model has optimal performance on a new task with limited samples after a small number of gradient updates. In our work, we consider MAML~\cite{finn2017model} and Meta-SGD~\cite{li2017meta} which are model- and task-agnostic. These optimization-based meta-learning algorithms promise to extract and propagate transferable representations of prior tasks. As a result, if we regard each task as learning to predict user preference for selecting recommendation models, each user will not only receive personalized model selection suggestions but also benefit from the choices of other users who have similar latent features.

\subsection{Model Selection for Recommender Systems}
In recommender systems, there is no single-best model that gives the optimal results for each user due to the heterogeneous data distributions among users. This means that the recommendation quality largely varies between different users~\cite{ekstrand2014user} and some users may receive unsatisfactory recommendations. One way to solve this problem is to give users the right to choose or switch the recommenders. As a result, explicit feedback can be collected from a subset of users to generate initial states for new users~\cite{ekstrand2015letting,dooms2013dynamic,resnick1994grouplens:}. Another solution is a hybrid recommender system \cite{burke2002hybrid}, which combines multiple models to form a complete recommender.
This type of recommender can blend the strengths of different recommendation models.
There are two types of methods to hybridize recommenders. One is to make a soft selection choice, that is, to compute a linear combination of individual scoring functions of different recommenders. A well-known work is feature-weighted-linear-stacking (FWLS) \cite{sill2009feature-weighted} which learns the coefficients of model predictions with linear regression. The other line of research is to make a hard decision to select the best individual model for the entire dataset \cite{cunha2016selecting, cunha2018metalearning}, for each user \cite{ekstrand2012when} or for each sample \cite{collins2018one-at-a-time:}. However, most of the works mentioned above are limited to collaborative filtering algorithms and require manually defined meta-features which is very time-consuming. Besides, despite the considerable performance improvement, methods like FWLS mainly focus on sample-level optimization which lacks interpretability about why some models work well for particular users, but not for others. In contrast, our proposed \ms can be trained end-to-end without extra meta-features. To our knowledge, our proposed framework is the first to explore the model selection problem for CTR Prediction, rather than collaborative filtering. We also provide an insight into which level to optimize in model selection by conducting extensive experiments for sample-level and user-level model selection.

\begin{figure}[t]
    \centering
    \includegraphics[width=0.45\textwidth]{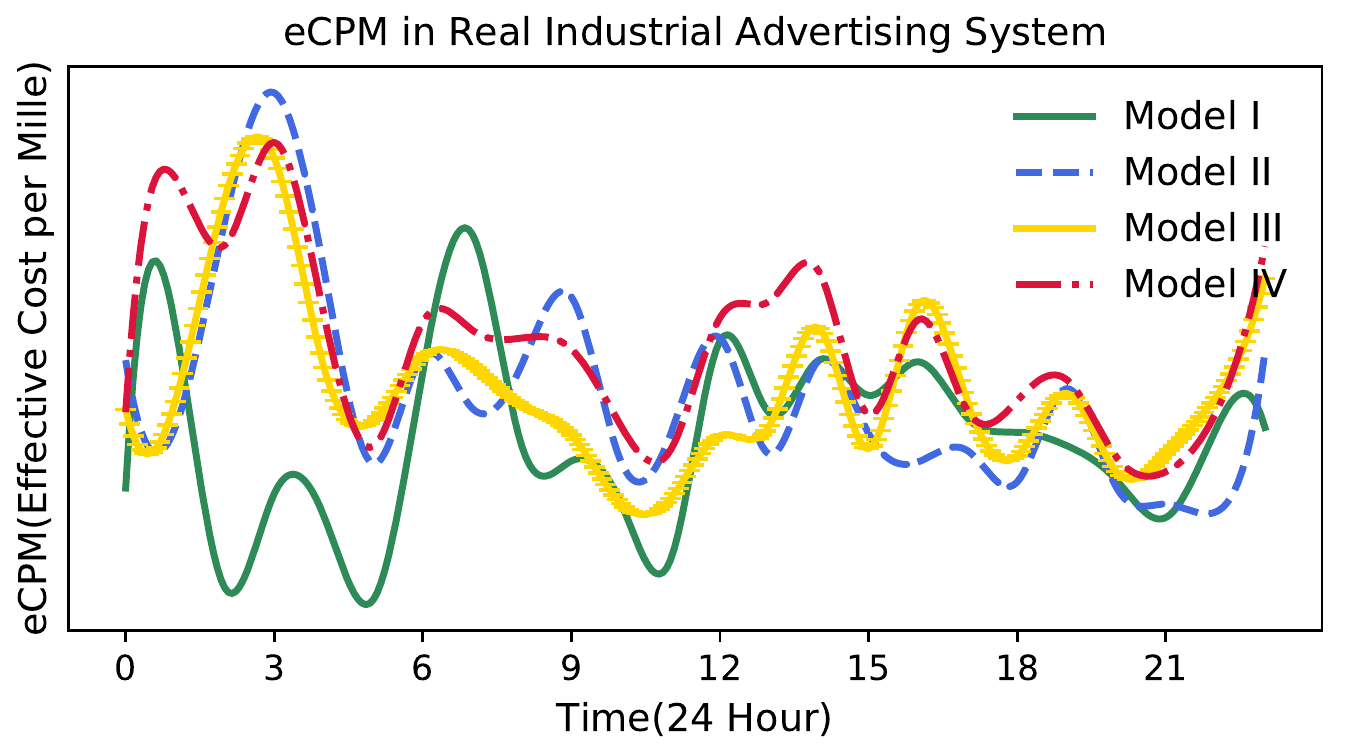}
    \vspace{-10pt}
    \caption{The performances of four models in one day.}
    \label{fig:online_eCPM}
    \vspace{-25pt}
\end{figure}

\subsection{CTR Prediction}
 Click-through rate (CTR) prediction is an important task in cost-per-click (CPC) advertising system. Model architectures for CTR prediction have evolved from shallow to deep. As a simple but effective model, Logistic Regression has been widely used in the advertising industry~\cite{chapelle2015simple, mcmahan2013ad}. Considering feature conjunction, Rendel presented Factorization Machines (FMs) which learn the weight of feature conjunction by factorizing it into a product of two latent vectors~\cite{rendle2010factorization}. As a variant of FM, Field-aware Factorization Machines (FFM) has been proven to be effective in some CTR prediction competitions~\cite{juan2016field,juan2017field}. To capture higher-order feature interactions, model architectures based on deep networks have been subsequently developed. Examples include Deep Crossing ~\cite{shan2016deep}, Wide \& Deep ~\cite{cheng2016wide}, PNN ~\cite{qu2016product-based}, DeepFM ~\cite{guo2017deepfm} and DIN ~\cite{zhou2018deep}. 

\section{Performance analysis}
In this section, we firstly present our observations about the varying online performance of recommendation models in a real industrial advertising system. Next, we conduct some pilot experiments to quantify this phenomenon with two public datasets. 

\subsection{Model Performance in Online Test}
In order to compare the performances of different models, we implement four state-of-the-art CTR prediction models, including shallow models and deep models. Then we deploy these models in a large-scale advertising system to verify the varying performances of them through online A/B test.

\textbf{Experimental Setting.} Users have been split into four groups, each of which contains at least one million users. Each user group receives recommendations from one of the four models. Our advertising system uses first price ranking approach, which means the candidate ads are ranked by bid*pCTR and displayed with the descending order. The bid is offered by the advertisers and the pCTR is generated by our CTR prediction model. The effective cost per mille (eCPM) is used as the evaluation metric:
\begin{equation}
    eCPM = \frac{Total Ads Income}{Total Ads Impressions} \times 1000.
    \label{con:inventoryflow}
\end{equation}

\textbf{Observations in Online Experiments.} We present the trends of eCPM values for four models within 24 hours in Figure~\ref{fig:online_eCPM}. Because of the commercial confidential, the absolute values of eCPM are hidden. We see that during the online A/B test, there is no single model which can achieve all-time best performance. For example, in general, Model I and Model III perform poorly during the day. However, Model I and Model III achieve leading performances from 7 a.m. to 8 a.m. and from 5 p.m. to 6 p.m. respectively. We also notice that although Model IV performs best on average, its eCPM is lower than that of some other models in particular time periods.

\begin{table}[t]
	\centering
	\caption{User proportion of different models.}
	\vspace{-10pt}
	\label{table:user_dictribution_public}
	\begin{tabular}{c|c c c c}
    	\toprule[1pt]
    		\textbf{Dataset} & \textbf{LR} & \textbf{FM} & \textbf{FFM}& \textbf{DeepFM}\\
    		\hline
    		Movielens-1m & 21.37$\%$ & 18.49$\%$ & 20.11$\%$ & 40.03$\%$ \\
    		Amazon-Electronics & 13.73$\%$ & 13.61$\%$ & 20.08$\%$ & 52.58$\%$ \\
    	\bottomrule[1pt]
	\end{tabular}
	\vspace{-15pt}
\end{table}

\subsection{Model Performance on Public Datasets}
We conducted some pilot experiments on MovieLens ~\cite{Harper2015The} and Amazon Review ~\cite{he2016ups} datasets to quantify the varying performance of models over different users. We consider four models (LR, FM~\cite{rendle2010factorization}, FFM~\cite{juan2016field} and DeepFM~\cite{guo2017deepfm}). We select the best model for each user by comparing the LogLoss.

As shown in Table \ref{table:user_dictribution_public}, in general, DeepFM performs better than other models: It is the best model for nearly 40$\%$ users in MovieLens, and the best for more than 52$\%$ users in Amazon. Although FM is the least popular model for both datasets, there are still 18.49$\%$ users in MoviesLens and 13.61$\%$ users in Amazon choosing FM.
\section{Methodology}
\label{sec:method}
In this section, we elaborate technical details for our proposed model selection framework \ms.
Suppose there is a set $U$ of users, where each user $u\in U$ has a dataset $D^u$ available for model training. A data point $(x, y)\in D^u$ consists of feature $x$ and label $y$. Note that our proposed framework provides a general training protocol for recommendation models, and is independent of specific model structure and data format.

\subsection{The \ms Framework}
The framework \ms consists of two major modules: the base models module and the model selection module. Next we describe the details of the workflow.

\textbf{Base models module.} A base model $\calM$ refers to a parameterized recommendation model, such as LR or DeepFM. A model $\calM$ with parameter $\theta$ is denoted by $\calM(\cdot;\theta)$, such that given feature $x$, the model outputs $\calM(x;\theta)$ as the prediction for the ground truth label $y$. Suppose in the base models module there are $K$ models $\calM_1, \calM_2, \ldots, \calM_K$, where $\calM_k$ is parameterized by $\theta_k$. Note that the $\calM_k$'s could have different structures, and hence contain distinct parameters $\theta_k$'s. In general the module allows different input features for different base models, while in what follows we assume all models have the same input form for ease of exposition.

\textbf{Model selection module.} This module contains a model selector $\calS$ that operates on top of the base models module. The model selector $\calS$ takes as input the data feature $x$ and outputs of base models $\calM(x;\theta):=(\calM_1(x;\theta_1), \calM_2(x;\theta_2), \ldots, \calM_K(x;\theta_K))$ where $\theta:=(\theta_1, \theta_2, \ldots, \theta_K)$, and outputs a distribution on base models. Suppose $\calS$ is parameterized by $\varphi$, the selection result is thus $\calS(x, \calM(x;\theta); \varphi)$. In practice, $\calS$ can be a multilayer perceptron (MLP) that takes $x$ only as input (without $\calM(x;\theta)$) and generates a distribution $\lambda=\calS(x;\varphi)$ over the base models, and the final prediction is the corresponding weighted average $\langle\lambda, \calM(x;\theta)\rangle$.

\subsection{Meta-training \ms}
The key ingredient that differentiates \ms with previous model selection approaches is that we use meta-learning to learn the model selector $\calS$, as shown in Algorithm~\ref{algo:ms_v1}. Our algorithm extends MAML into the \ms framework. The original MAML is applied to a single prediction model, while in our case MAML is used to jointly learn the model selector and base models.

\IncMargin{0.1em}
\begin{algorithm}[t]
	\SetAlgoLined
	
	\caption{\ms}
	\label{algo:ms_v1}
	
	\KwData{Training set $D^u$ for user $u\in U$}
	Initialize $\theta_k$ for $\calM_k$ with $k\in \{1,\ldots,K\}$, and $\varphi$ for $\calS$\;
	
	Denote $\theta=(\theta_1, \ldots, \theta_K)$ and $\lambda = (\lambda_1, \ldots, \lambda_K)$\;
	
	(Optional) Pretrain $\theta$ using $\bigcup_{u\in U} D^u$\;
	
	\ForEach {\emph{episode $t=1,2,...$}}{
		
		Sample a set $U_t$ of $m$ users from $U$ \;
		
		\ForEach {\emph{user $u\in U_t$}}{
			Sample $D^u_S$ and $D^u_Q$ from $D^u$\;
			
			\ForEach{$(x,y)\in D^u_S$}{
			    $\lambda \leftarrow \calS(x; \varphi)$\;
			    
				$p(x;\theta,\varphi) \leftarrow \sum_{k=1}^{K} \lambda_k \calM_k(x;\theta_k)$\;
			}
			
			$\mathcal{L}_{D^u_S}(\theta, \varphi) \leftarrow \frac{1}{|D^u_S|} \sum_{(x, y) \in D^u_S} \ell(p(x;\theta,\varphi), y)$\;
			
			$(\theta^u, \varphi^u) \leftarrow (\theta, \varphi) - \alpha \nabla_{\theta,\varphi} \mathcal{L}_{D^u_S}(\theta, \varphi)$\;
			
			\ForEach{$(x,y)\in D^u_Q$}{
			    $\lambda \leftarrow \calS(x; \varphi^u)$\;
			    
				$p(x;\theta^u,\varphi^u) \leftarrow \sum_{k=1}^{K} \lambda_k \calM_k(x;\theta^u_k)$\;
			}
			
			$\mathcal{L}_{D^u_Q}(\theta^u, \varphi^u) \leftarrow \frac{1}{|D^u_Q|} \sum_{(x, y) \in D^u_Q} \ell(p(x;\theta^u,\varphi^u), y)$\;
		}
		$(\theta, \varphi) \leftarrow (\theta, \varphi) - \beta \cdot \frac{1}{m}\sum_{u\in U_t} \nabla_{\theta, \varphi} \mathcal{L}_{D^u_Q}(\theta^u, \varphi^u)$\;
	}
\end{algorithm}

\textbf{Episodic Meta-training.}
The meta-training process proceeds in an episodic manner. In each episode, a batch of users are sampled as tasks from a large training population (line 5). For each user $u$, a \emph{support} set $D^u_S$ and a \emph{query} set $D^u_Q$ are sampled from $D^u$, which are considered as ``training'' and ``test'' sets in the task corresponding to user $u$, respectively (line 7). We adopt the common practice in meta-learning literature that guarantees no intersection between $D^u_S$ and $D^u_Q$ to improve generalization capacity. After an in-task adaptation procedure is performed for each task (lines 8--18), at the end of an episode, the \emph{initialization} $\varphi$ for the model selector and  $\theta$ for base models are updated according to the loss signal received from in-task adaptation (line 20).
Here the initialization is maintained and will be adapted to new user when deployed.
Next we describe the in-task adaptation procedure.

\textbf{In-task Adaptation.}
Given the currently maintained parameters $\theta$ and $\varphi$, the \ms first iterates the support set $D^u_S$ to generate a per-item distribution $\lambda$ on base models (line 9), and then get a final prediction $p(x;\theta,\varphi)$ which is a convex combination of outputs $\calM(x;\theta)$ (line 10). The training loss $\mathcal{L}_{D^u_S}(\theta^u, \varphi)$ is computed by averaging $\ell(p(x;\theta,\varphi), y)$ over data points in $D^u_S$ (line 12), where $\ell$ is a pre-defined loss function. In this work we focus on CTR prediction problems and use LogLoss as the loss function:
\begin{align}
    \ell(p(x;\theta,\varphi), y) = -y \log p(x;\theta,\varphi) - (1-y) \log (1-p(x;\theta,\varphi)),
\end{align}
where $y\in \{0,1\}$ indicates if the data point is a positive sample.
Then a gradient update step is performed to parameters of the base models and model selector, leading to a new set of parameters $\theta^u$ and $\varphi^u$ adapted to the specific task (line 13).
The test loss $\mathcal{L}_{D^u_Q}(\theta^u, \varphi^u)$ is then computed on the query set in a similar way as computing training loss, using the updated parameters of base models and model selector instead (lines 14--18).
Note that by keeping the path of in-task adaptation (from $(\theta,\varphi)$ to $(\theta^u,\varphi^u)$), the test loss $\mathcal{L}_{D^u_Q}(\theta^u, \varphi^u)$ can be expressed as a function of $\theta$ and $\varphi$, which is passed to the outer loop for updating $\theta$ and $\varphi$ using gradient descent methods such as SGD or Adam.

\textbf{Jointly Meta-training $\theta$ and $\varphi$.}
We further note that $\theta$ and $\varphi$ are updated together in the outer loop (line 20) that serve as initialization for the base models and model selector, respectively.
The parameters are updated to adapt to each user (line 13). This step is crucial for \ms to operate at the user level, i.e., to execute user-level model selection via base models and model selector modules adaptive to specific users.
The episodic meta-learning procedure plays an important role to obtain learnable initialization for \ms to enable fast adaptation on users.
The objective of meta-training can be formulated as follows:
\begin{align}
    \textstyle \min\nolimits_{\theta,\varphi} \mathbb{E}_{u\in U} \left[\mathcal{L}_{D^u_{Q}} \left((\theta,\varphi) - \alpha \nabla_{\theta,\varphi}\mathcal{L}_{D^u_S}(\theta,\varphi)\right)\right].
\end{align}

\textbf{Learning Inner Learning Rate $\alpha$.}
The inner learning rate $\alpha$, which is often a hyper-parameter in normal model training protocols, can also be learned in meta-learning approaches by considering the test loss $\mathcal{L}_{D^u_Q}(\theta^u, \varphi^u)$ as a function of $\alpha$ as well. Li et al.~\cite{li2017meta} showed that learning per-parameter inner learning rate $\alpha$ (a vector of same length as $\theta$) achieves consistent improvement over MAML for regression and image classification. Algorithm~\ref{algo:ms_v1} can be slightly modified accordingly: in line 13, the inner update step becomes:
\begin{align}
    \textstyle (\theta^u, \varphi^u) \leftarrow (\theta, \varphi) - \alpha\circ \nabla_{\theta,\varphi} \mathcal{L}_{D^u_S}(\theta, \varphi),
\end{align}
where $\circ$ denotes Hadamard product. Considering $\theta^u$, $\varphi^u$ as a function of $\alpha$, the outer update step in line 20 becomes:
\begin{align}
    \textstyle (\theta, \varphi, \alpha) \leftarrow (\theta, \varphi, \alpha) - \beta \cdot \frac{1}{m}\sum_{u\in U_t} \nabla_{\theta, \varphi, \alpha} \mathcal{L}_{D^u_Q}(\theta^u, \varphi^u),
\end{align}
where gradients flow to $\alpha$ through $\theta^u$ and $\varphi^u$. The objective function can be accordingly written as:
\begin{align}
    \textstyle \min\nolimits_{\theta,\varphi,\alpha} \mathbb{E}_{u\in U} \left[\mathcal{L}_{D^u_{Q}} \left((\theta,\varphi) - \alpha\circ \nabla_{\theta,\varphi}\mathcal{L}_{D^u_S}(\theta,\varphi)\right)\right].
\end{align}
In practice we find that learning a vector $\alpha$ could significantly boost the performance of \ms for recommendation tasks.

\textbf{Meta-testing/Deployment.}
Meta-testing \ms on new tasks follows the same in-task adaptation procedure as in meta-training (lines 7--17), after which evaluation metrics are computed such as AUC and LogLoss.
A separate group of meta-testing users (with no intersection with meta-training users) may be considered to justify the generalization capacity of meta-learning on new tasks.

\textbf{Simplifying \ms.}
We propose a simplified version of meta-training for \ms, where no in-task adaptation for base models is required. The base models are pre-trained before meta-training and then fixed. The model selector is trained episodically. We note that this procedure is also in the meta-learning paradigm since $\varphi$ is updated using user-wise mini-batches, where for each user $u$ the distribution $\lambda$ is generated using a support set $D^u_S$, and evaluated by computing test loss on a separate query set $D^u_Q$. This enables \ms to learn at user level and generalize to new users efficiently. At meta-testing phase, base models as well as the model selector are fixed, and the training set is simply used for the model selector to generate a distribution over base models.
The simplified \ms may be of particular interest in practical recommender systems where in-task adaptation is restricted due to computation and time costs, such as news recommendation for mobile users using on-device models.

\section{experiment}
\label{sec:exp}

In this section, we evaluate the empirical performance of the proposed method, and mainly focus on CTR Prediction tasks where the prediction quality plays a very important role and has a direct impact on the business revenue. We experiment with two public datasets and a real-world production dataset. The statistics of the selected datasets are summarized in Table~\ref{table:dataset}. We raise and try to address two major research questions:
\noindent
\begin{itemize}
    \item \textbf{RQ1}: Can model selection help CTR Prediction? 
    \item \textbf{RQ2}: What benefits could \ms bring to personalized model selection?
\end{itemize}

\begin{table}[t]
	\centering
	\caption{Statistics of selected datasets.}
    \vspace{-10pt}

	\label{table:dataset}
	\begin{tabular}{c |c c c c}
	\toprule[1pt]
		\textbf{Dataset} & \textbf{Users} & \textbf{Items} & \textbf{Samples} & \textbf{Features}\\
		\hline
		Movielens-1m & 6,040 & 3,952 & 1,000,209 & 14,025\\
		Amazon-Electronics & 192,403 & 63,001 & 1,689,188 & 319,687\\
		Production Dataset & 7,684 & 2,420 & 3,333,246 & 11,860\\
	\bottomrule[1pt]
	\end{tabular}
    \vspace{-15pt}
\end{table}

\subsection{Datasets}
\textbf{Movielens-1m.} Movielens-1m~\cite{Harper2015The} contains 1 million movie ratings from 6040 users and each user has at least 20 ratings. We regard 5-star and 4-star ratings as positive feedbacks and label them with 1, and label the rest with 0. We select the following features: user\_id, age, gender, occupation, user\_history\_genre, user\_history\_movie, movie\_id, movie\_genre, day of week and season.

\textbf{Amazon-Electronics.} Amazon Review Dataset \cite{he2016ups} contains user reviews and metadata from Amazon and has been widely used for product recommendation. We select a subset called Amazon-Electronics from the collection and shape it into a binary classification problem like Movielens-1m. Following \cite{he2015vbpr:}, we use the 5-core setting to retain users with at least 5 ratings. The selected features include user\_id, item\_id,item\_category, season, user\_history\_item (including 5 products recently rated), user\_history\_categories.

\textbf{Production Dataset.}
To demonstrate the effectiveness of our proposed methods on real-world application with natural data distribution over users, we also evaluate our methods on a large production dataset from an industrial recommendation task. Our goal is to predict the probability that a user will click on the recommended mobile services based on his or her history behavior. In this dataset, each user has at least 203 history records.

\subsection{Baselines}
We compare the proposed methods with two kinds of competitors: single models and hybrid recommenders with model selectors.

\begin{table*}[t]
	\centering
	\caption{AUC and LogLoss Results.}
    \vspace{-5pt}
	\label{table:AUC}
	\begin{tabular}{c|c c c|c c c|c c c}
	\toprule[1pt]
		\textbf{Model} & \multicolumn{3}{c|}{\textbf{Movielens.}} & \multicolumn{3}{c|}{\textbf{Amazon.}} & \multicolumn{3}{c}{\textbf{Production.}}\\
		\cline{2-10}
		& AUC & LogLoss & RelaImpr & AUC & LogLoss & RelaImpr & AUC & LogLoss & RelaImpr\\
		\hline
		LR & 0.7914 & 0.55112 & -1.45\% & 0.6981 & 0.46374 & -6.29\% & 0.7813 & 0.54011 &  -29.83\%\\
		FM & 0.7928 & 0.54917 & -0.98\% & 0.6953 & 0.46242 & -7.62\% & 0.8821 & 0.42618 & -4.69\%\\
		FFM & 0.7936 & 0.54826 & -0.71\% & 0.7114 & 0.45216 & 0.00\% & 0.8850 & 0.42469 & -3.97\%\\
	    DeepFM & 0.7957 & 0.54672 & 0.00\% &0.7101 & 0.45696 & -0.61\% & 0.9009  & 0.39215& 0.00\% \\
		\hline
		Perfect Sample-level Selector & 0.9008 & 0.41079 & 35.54\% &0.8411 & 0.37088 & 61.35\% & 0.9710 & 0.26043 & 17.49\%\\
		Perfect User-level Selector & 0.8187 & 0.51829 & 7.78\% &0.8135 & 0.38999 & 48.30\% & 0.9051 & 0.37835 & 1.05\%\\	    
		\hline
		Sample-level Selector & 0.7963 & 0.54482 & 0.20\% & 0.7121 & 0.45152 & 0.33\% & 0.9011 & 0.39137 & 0.05\%\\
		User-level Selector & 0.7999 & 0.54058 & 1.42\% & 0.7124 & 0.45117 & 0.47\% & 0.9013 & 0.39109 & 0.10\%\\
		\ms-Simplified & 0.8036 & 0.53550 & 2.67\% &0.7134 & 0.45044 & 0.95\% & 0.9022 & 0.39095 & 0.32\%\\
		\ms & \textbf{0.8047} & \textbf{0.53531} & \textbf{3.04\%} & \textbf{0.7141} & \textbf{0.44996} & \textbf{1.28\%} &  \textbf{0.9023} & \textbf{0.39036} & \textbf{0.35\%}\\
	\bottomrule[1pt]
	\end{tabular}
	\vspace{-6pt}
\end{table*}

\textbf{Single Models.} We consider three types of model architectures, including linear (LR), low rank (FM~\cite{rendle2010factorization} and FFM~\cite{juan2016field}) and deep models (DeepFM~\cite{guo2017deepfm}). The latent dimension of FM and FFM is set to 10. The field numbers of FFM for Movielens, Amazon and Production are 22, 18 and 8 respectively. For DeepFM, the dropout setting is 0.9. The network structures for Movielens, Amazon and production datasets are 256-256-256, 400-400-400 and 400-400-400 respectively. We use ReLU as the activation function.

\textbf{Sample-level Selector and User-level Selector.} These two methods are used as model selection competitors. They are designed to predict the model probability distribution for each sample and for each user. 80\% local data of each user is used for training and the rest for testing.
Then the local data of all users is collected to generate the whole training and testing data.
While training, 75\% training data is firstly used to train four CTR prediction models in a mini-batch way~\cite{li2014efficient}.
The batch size is set to 1000.
Then the pretrained recommenders predict the CTR values and Logloss for the remaining training data. For the two baselines, we give sample-level and user-level labels from 0-3 by comparing LogLoss respectively. As for additional meta-features used to train a 400-400-400 MLP classifier, we consider the CTR prediction values of the four recommenders. While testing, the final prediction for each sample or each user is the weighted average of the predicted values of the individual models.

\subsection{Settings and Evaluation Metrics}
For \ms, the division of user local data is the same as the division for sample-level MLP selector. During meta-training process, the training data of each user is further divided into 75\% support set and 25\% query set. During meta-testing phase, the model selector and base models are firstly fine-tuned before evaluating on the testing data. The performance metrics used in our experiments are AUC (Area under ROC), LogLoss and RelaImpr. RelaImpr is calculated as follows:
\begin{equation}
    RelaImpr=\left(\frac{AUC(to\;be\;compared)-0.5}{AUC(\\single\;best\;model)-0.5}-1\right)\times100\%.
\end{equation}
For pre-training of CTR models, we use FTRL optimizer~\cite{mcmahan2010adaptive} for LR and Adam optimizer \cite{kingma2014adam} for FM, FFM and DeepFM. The mini-batch size is 1000. For \ms, we use Meta-SGD \cite{li2017meta} to adaptively learn the inner learning rate $\alpha$. The initial value of inner learning rate $\alpha$ for Movielens, Amazon and Production dataset is 0.001, 0.0001, 0.001. The outer learning rate $\beta$ is set to $1/10$ of $\alpha$. In each episode of meta-training, the numbers of active users are 10. We use a 200-200-200 MLP as the model selector.

\subsection{Performance of Model Selection}
\textbf{RQ1: Overall Performance Comparison.}
To investigate RQ1, we study the performance of baselines and \ms on three datasets, the results are summarized in Table \ref{table:AUC}. To explore the potential and limit of model selection approaches, we compute the upper bound through two perfect model selectors: (1) perfect sample-level selector which chooses the best model for each sample; (2) perfect user-level selector which chooses the best model for each user. First, comparing single model baselines with hybrid recommender with model selection, we see that all model selection methods achieve a considerable improvement in terms of AUC and Logloss. This result is highly encouraging, indicating the effectiveness of model selection methods. Second, comparing the sample-level selectors with the user-level selectors, we find that perfect sample-level model selector is expected to achieve greater improvements than perfect user-level selector.
However, in the last four rows of Table~\ref{table:AUC}, we show the performance of actual selectors and observe that user-level selectors achieve higher AUC and lower Logloss, rather than the sample-level selector. This discovery implies that the differences between samples may be too subtle for the selector to be well fitted. In contrast, the latent characteristics of different users vary widely, which makes the \ms work well. Finally, we compare \ms and \ms-simplified, finding that the performance of the simplified version dropped slightly. This verifies our argument in Section \ref{sec:method} that the in-task adaptation could make model selection more user-specific.

\begin{figure}[t]
    \centering
    \includegraphics[width=0.38\textwidth]{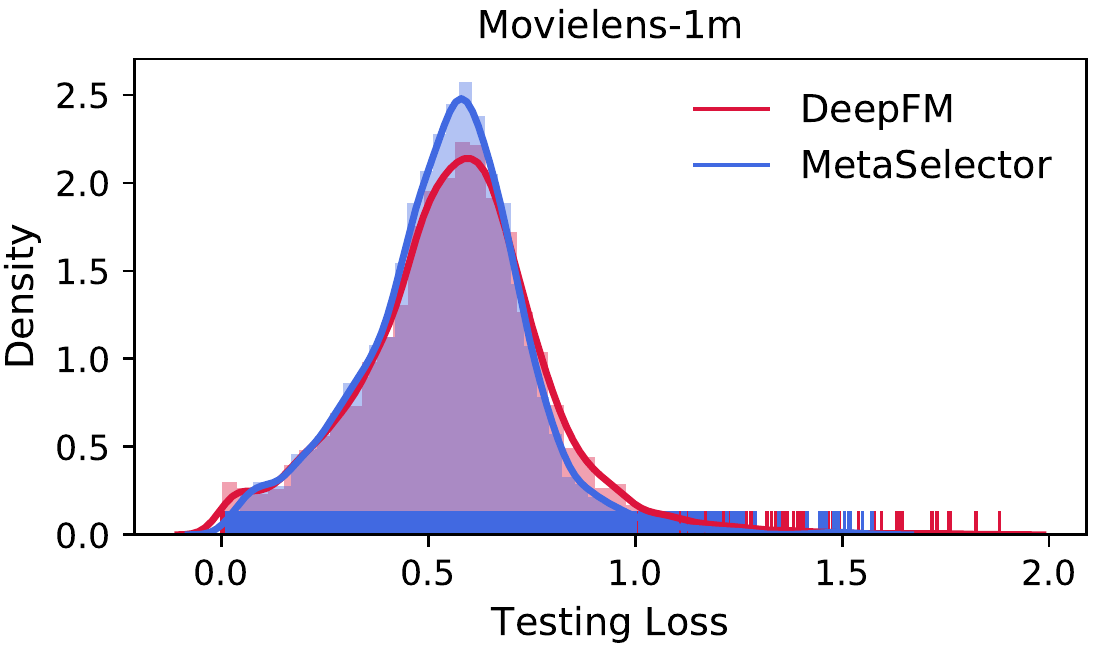}
    \vspace{-10pt}
    \caption{KDE for Movielens.}
    \vspace{-15pt}
    \label{fig:ml_kde}
\end{figure}

\textbf{RQ2: Performance Distribution Analysis.}
Despite the overall improvement, it is also worth studying RQ2: In what ways does \ms help model selection? To this end, we further investigate the testing loss distribution on all users with MovieLens-1m dataset. Figure \ref{fig:ml_kde} shows the kernel density estimation of \ms and DeepFM which is a strong single model baseline. We observe that \ms not only leads to lower mean LogLoss but also achieves more concentrated loss distribution with lower variance. This shows that \ms encourages a more fair loss distribution across users and is powerful to model heterogeneous users. The above observations verify the effectiveness of our proposed methods in terms of personalized model selection. 

\section{Conclusions}
In this work, we addressed the problem of model selection for recommender systems, motivated by the observation of varying performance of different models among users on public and private datasets.
We initiated the study of user-level model selection problems in recommendation from the meta-learning perspective, and proposed a new framework \ms to formulate a user-level model selection module.
We also ran extensive experiments on both public and private production datasets, showing that \ms can improve the performance over single model baseline and sample-level selector. This shows the potential of \ms in real-world recommender systems.



\bibliographystyle{ACM-Reference-Format}
\bibliography{meta-learning,rec}


\end{document}